\def\ie{{\frenchspacing\it i.e.}}
\def\eg{{\frenchspacing\it e.g.}}
\def\etc{{\frenchspacing\it etc.}}

\def\rf#1;#2;#3;#4;#5 {\par#1, {\it #3} {\bf #4}, #5 (#2). \par}

\def\beq#1{\begin{equation}\label{#1}}
\def\eeq{\end{equation}}
\def\beqa#1{\begin{eqnarray}\label{#1}}
\def\eeqa{\end{eqnarray}}

\def\ignore#1{}

\def\R{{\bf R}}
\def\vr{{\bf r}}
\def\vp{{\bf p}}
\def\opr{\widehat{\vr}}
\def\op{\widehat{\vp}}

\def\spose#1{\hbox to 0pt{#1\hss}}
\def\simlt{\mathrel{\spose{\lower 3pt\hbox{$\mathchar"218$}}
     \raise 2.0pt\hbox{$\mathchar"13C$}}}
\def\simgt{\mathrel{\spose{\lower 3pt\hbox{$\mathchar"218$}}
     \raise 2.0pt\hbox{$\mathchar"13E$}}}
\def\simpropto{\mathrel{\spose{\lower 3pt\hbox{$\mathchar"218$}}
     \raise 2.0pt\hbox{$\propto$}}}

\def\ed{\end{document}}
\def\bl{\vskip0.423truecm}
\def\ind{\noindent\hskip1.8truecm}
\def\Section#1{{\raggedright\bl\bl\goodbreak{\noindent\bf#1}\bl}}
\def\Subsection#1{{\raggedright\vskip0.564truecm\goodbreak{\noindent\bf#1}\bl}}
\def\Sectionn#1#2#3{\Section{\hskip-0.25cm\begin{tabular}{ll}#1&#2\\&#3\end{tabular}}}
\def\Subsectionn#1#2#3{\Subsection{\hskip-0.25cm\begin{tabular}{ll}#1&#2\\&#3\end{tabular}}}

\documentstyle[11pt,smartcite]{article}

\begin{document}
\baselineskip=0.423truecm
\def\basel{\normalsize\baselineskip=0.423truecm}

\pagestyle{empty}


\noindent
{\it Foundations of Physics Letters, Vol. 9, No. 1, 1996, pages 25-42}
\vskip0.9truecm

\noindent
{\bf 
DOES THE UNIVERSE IN FACT CONTAIN\\
ALMOST NO INFORMATION?
}
  
\bl\bl\bl
\ind
Max Tegmark

\bl\ind 
{\it Max-Planck-Institut f\"ur Physik}

\ind 
{\it F\"ohringer Ring 6}

\ind   
{\it D-80805 M\"unchen, Germany}

\ind   
{\it max@mppmu.mpg.de}

\bl\bl\bl\ind
Received June 2, 1995; revised November 10, 1995

\bl\bl

\noindent
At first sight, an accurate description of the state of the 
universe appears to require a mind-bogglingly large and 
perhaps even infinite amount of information, even if we 
restrict our attention to a small subsystem such as a 
rabbit.
In this paper, it is suggested that most of this information is 
merely apparent, as seen from our subjective viewpoints, 
and that the algorithmic information content of the 
universe as a whole is close to zero.
It is argued that if the Schr\"odinger equation is
universally valid, then decoherence together with
the standard chaotic behavior of
certain non-linear systems will 
make the universe appear extremely complex
to any self-aware subsets that happen to inhabit it
now, even if it was in a quite simple state shortly 
after the big bang.
For instance, 
gravitational instability would amplify the microscopic 
primordial density fluctuations that are required by the 
Heisenberg uncertainty principle into quite macroscopic inhomogeneities,
forcing the current wavefunction of the universe to contain such Byzantine
superpositions as our planet being in many macroscopically 
different places at once.  
Since decoherence bars us from experiencing more than one macroscopic reality, 
we would see seemingly complex constellations of stars {\etc}, 
even if the initial 
wavefunction of the universe was perfectly homogeneous and isotropic.
\bl


\vskip1.3truecm
\noindent
{\footnotesize
Online at {\it http://www.mpa-garching.mpg.de/$\sim$max/nihilo.html} (faster 
from Europe)\\
and at {\it http://astro.berkeley.edu/$\sim$max/nihilo.html} (faster from 
the US).
}

\clearpage

\Section{1. INTRODUCTION}

\noindent
One of the most striking features of the universe we inhabit is its
complexity. When going swimming after setting up an
over-night hydrodynamics batch job on the computer, it is easy to
start marveling over the system of 
currents, waves and vortices in the pool, and ask oneself 
silly questions like
``how on earth can 
nature calculate all this in real time?".
Quite apart from the fact that many of the non-linear 
partial differential equations that  
nature appears to integrate with ease are numerically ill-posed
as initial value problems, exhibiting chaotic behavior, 
the sheer number of bytes required to store the initial data for a system
such as Niagara falls at a reasonable resolution are mind-boggling.
Indeed, if we were to take classical
General Relativity seriously, then space is a continuum locally 
isomorphic to $\R^3$, which means that to describe the location
of something we need to specify three real numbers. Of course, this
already constitutes an infinite about of information
--- to specify a generic real number completely would
be equivalent to specifying an infinite number of 
seemingly random decimals.
Likewise, to specify a quantity such as the electric field ${\bf E}$ 
at even a single point would involve specifying real numbers, {\ie},
an infinite amount of information.
This has disturbed many authors in the past, and has
been one of the motivations for attempts to
replace continuum physics by some discrete alternative,
{\eg} \cite{Yukawa,Yamamoto,Plokhotnikov}.

In this paper, we will instead take the viewpoint of the old 
phrase ``it's just a figment of your imagination", and argue that 
continuum physics can be maintained without involving infinite 
quantities of information, and indeed without invoking any 
new physics whatsoever.
The argument will basically go as follows.
\begin{itemize}
\item
The wave function of the universe shortly after the big
bang had some quite simple form, which can be described with very 
little algorithmic information.
\item
By the Heisenberg uncertainty principle, this initial state
involved micro-superpositions, microscopic quantum 
fluctuations in the various fields.
\item
The ensuing time-evolution involved non-linear elements
that exhibited chaotic behavior (such as the well-known gravitational
instability that is held responsible for the formation of cosmic
large-scale structure), which amplified these micro-superpositions 
into macro-superpositions.
\item
In the no collapse version of quantum mechanics,
the current wavefunction of the universe is
thus a superposition of a large number of states that
are macroscopically extremely different (Earth forms here, 
Earth forms one meter further north, {\it etc}).
\item
Since macroscopic objects inevitably interact with their surroundings,
the well-known effects of decoherence will
prevent self-aware subsets of the universe (such as us) from 
perceiving such macro-superpositions.
\item
The net result is that although the wavefunction of the universe 
contains almost no algorithmic information (we can specify it 
by simply giving the initial wave function and the Hamiltonian),
and will retain for instance translational and rotational symmetry 
from the initial data, we will experience an asymmetric 
universe that appears extremely complex.
\end{itemize}
In other words, the ensemble of all branches in the 
current macro-superposition
is simple to describe, but to describe any one element of the 
ensemble (because of decoherence, that is all we will observe),
requires an enormous amount of information. This is much like
the fact that one needs very little information
to describe a real-valued random variable with a simple probability
distribution, but an infinite amount of information to describe
any one realization of it.
We will now discuss the elements of this argument in somewhat 
greater detail.

\Sectionn{2.}{ALGORITHMIC INFORMATION AND}{INITIAL CONDITIONS}

\noindent
The {\it algorithmic information content} \citerange{Solomonoff}{Caves}
(also known as the {\it information content}, 
the {\it algorithmic complexity} or {\it algorithmic randomness}) 
of a bit string $s$ is defined as the length (in bits) of the shortest 
program for a universal computer that produces $s$ as output.
For instance, the information content of a string of $10^9$ random
bits would be about $10^9$ bits, whereas that of a string of $10^9$ 
zeroes would be quite small.
Subtle details such as assumed capabilities of the 
``universal computer" are discussed in detail in the literature, and 
are irrelevant to the present discussion. 
The above definition is straightforward to extend to wider classes of
objects than bit strings. Integers are of course bit strings when written in
binary form, and the definition of the information content of a real number
is analogous. 
The algorithmic information content of a wavefunction
$\psi$ is customarily (\eg, \cite{ZurekAlgo2}) defined as the 
length of the shortest computer program 
that given any point $\vr$ in the configuration space and 
a prescribed accuracy $\epsilon$ will evaluate 
$\psi(\vr)$ to this accuracy. 
Similarly, the information definition is readily widened to encompass 
the other
mathematical objects that will recur in this paper.

As an example, consider an image of the Mandelbrot set.
We can equivalently think of it as a scalar field in the complex plane, whose
value determines the color of the image at each point (say black or white). 
To the lay observer, it might 
appear to contain a vast and perhaps even infinite amount of 
information. However, since we can describe the Mandelbrot set
it in a single sentence,
simply as the set of points $c$ in the complex plane
for which the iteration $z\mapsto z^2+c$ will not diverge when we 
start with $z=0$, 
a short program can determine whether a 
given point belongs to the set or not,
and the algorithmic information content of the image is 
thus correspondingly small.

\Subsectionn{2.1.}{The Sense in Which the Whole is Less than}{a Sum of its Parts}

\noindent
In the case of the Mandelbrot set, the source of the 
discrepancy between apparent and actual (algorithmic) information
content was chaos, caused by nonlinear dynamics. 
We will return to this in {\frenchspacing Sec. 3}, as this is 
one of the two things that we will argue causes the abundance of 
apparent information around us.
The second source is related to quantum symmetry breaking,
and will be elaborated upon in {\frenchspacing Sec. 4}.
Loosely speaking, the apparent information content rises when
we restrict our attention to one particular element in an ensemble,
thus losing the symmetry and simplicity that was inherent in the totality
of all elements taken together.
A uniform probability distribution 
on the unit interval $[0,1]$ has a
very small algorithmic information content, since the distribution function
simply equals unity on this interval and vanishes outside of it.
Now let us draw one number from this distribution. 
It can be written out with infinitely many decimals, so
to describe it requires an infinite amount of algorithmic
information unless we are lucky enough to draw something
simple like a rational number, $1/\sqrt{2}$, $1/\pi$, {\etc},
and this rarely happens --- indeed, it is easy to prove
that the probability of drawing a number with a finite amount of algorithmic
information is zero.\footnote{\basel
Any computer program of finite length can 
be thought of as a rational number between 0 and 1 by interpreting its 
bits as binary decimals preceded by ``0.". 
Since the set of all rational numbers on $[0,1]$ has
Lebesgue measure zero, the result
follows.}

In short, any one realization contains more information
than the probability distribution itself.
Similarly, any one generic real number between zero and one 
contains more information than the totality of all of
them --- the algorithmic information content of a single 
number is infinite, whereas the entire set $[0,1]$ 
can be generated in lexicographical order by quite a 
simple computer program \cite{ZurekAlgo1}.

\Subsection{2.2. The Initial State of the Universe}

\noindent
In modern cosmology, the field of density fluctuations in
the universe (see {\eg} \cite{Kolb&Turner}) is usually
modeled as a random field $\delta(\vr)$, and it is often assumed
that this field was homogeneous, isotropic and Gaussian early on,
long before non-linear structures such as galaxies formed.
This means that the entire field (or rather the entire ensemble
of all fields) can be completely described by merely one  
single function of one variable, $\xi(r)$, the correlation function, 
which specifies the correlation of the field values at two points
separated by a distance $r$. 
If this function were given by a simple analytic expression, 
then this particular aspect of the initial 
data would thus have a very small algorithmic information content.
There is of course a catch here: the ensemble of all random fields
is merely a convenient statistical concept, and few cosmologists
would take this literally enough to believe that there are in fact
infinitely many universes with the same correlation function, us merely being
ignorant as to which one we happen to inhabit.
However, as will be argued in the two following sections, 
the no-collapse interpretation of quantum mechanics
provides a formalism which is in many ways similar to that of 
random fields, and in addition is supposed to describe actual
macroscopic realities that exist
in superposition.

A familiar example is the vacuum state of any field theory, which
could be interpreted as a superposition of an infinite number of
different field configurations. Although this is usually not  
useful when computing half-lives, cross sections {\etc}, 
such a vacuum can be characterized in a way quite analogous to
that in which random fields are defined, the only major difference
being that all probabilities are replaced by complex-valued amplitudes. 
If the vacuum is homogeneous and isotropic, the description
simplifies greatly just as in the above-mentioned random field
case. 
The key point is that the algorithmic information 
content of the vacuum
state of any field theory that we can define is finite, 
since by merely defining the field theory, we automatically define its 
vacuum. Although we know virtually nothing about quantum gravity and 
the state of the universe less than a Planck time after
the big bang, it does not appear 
particularly outrageous to postulate that the universe was once in
something resembling a vacuum state of a more complete field 
theory than we presently know of (incorporating quantum gravity),
or at least in some sort of state that would be equally 
simple to describe.
For the rest of our argument, we will simply assume that 
state of the universe at some early time exhibited so much symmetry that
it contained only a small and finite amount of algorithmic information.

\Section{3. CHAOS AND SYMMETRY BREAKING}

\noindent
Classical physics provides us with an abundance of familiar
examples where the apparent information content of a system rises although the
algorithmic information content does not, because of chaotic
behavior. Water approaching Niagara Falls and a window hit
by a football are two familiar examples. 
In both of these cases, the dynamics amplifies microscopic 
differences in the initial data into visible macroscopic differences.
Here we will center our discussion around
an example of gravitational instability.

\Subsection{3.1. The Classical Cosmological Density Field}

\noindent
Assuming that the laws of physics that govern the time-evolution
are known, the algorithmic information 
content of the final state is of course not
much greater than that of
the initial state, since the program that specifies
the final state can can simply contain the program that 
computes the initial state together with the appropriate numerical routines
(for integration of the partial differential equations of nature, say). 
The fact that the time required for the computation might be prohibitive
is irrelevant here, since algorithmic information refers only to the {\it size}
of the computer program, not to the time it takes to run it. 

As an example, let us assume that we know the field of cosmological 
density fluctuations at an early time as well as the physics
that describes its time evolution. Let us also assume that the initial
data are those of a standard cosmological structure formation scenario,
{\ie}, that the initial fluctuations are very small, corresponding 
to a nearly uniform density, with an appropriate power spectrum
(see, {\eg}, \cite{Kolb&Turner}).
To keep things simple, we can assume that all matter is collisionless
and that no other forces than classical gravity are at work (including
further elements of realism would of course make the system 
even more chaotic). This uniquely specifies the state of the matter
distribution today, and we know from numerical simulations that 
despite the simplicity of the initial data, this
state would be extremely complex, with highly non-linear clumping on a wide
range of scales.  
In other words, gravity alone creates apparent complexity of the
type we mentioned when discussing the Mandelbrot set. More precisely, 
gravitational dynamics
causes chaotic behavior, in the sense that 
a pair of very similar initial data sets could produce 
results that are macroscopically different  --- recall that 
gravitational dynamics exhibits chaotic behavior 
even in the much simpler case
of the three body problem.

\Subsection{3.2. The Quantum Case}

\noindent
To be able to appeal to the example of gravitational dynamics for our main 
argument, we need to take this one step further,
beyond classical physics.
If the initial state corresponded to {\it exactly} uniform density,
then so would the final state. Since the real world has a non-zero 
Planck constant, the actual initial state in our simple
gravity problem is of course described by a wavefunctional, $\psi$, 
that associates a complex amplitude to each configuration
of the field of density fluctuations, $\delta(\vr)$,
and the time-evolution is governed not by the classical equations
of motion for $\delta(\vr)$, but by the Schr\"odinger equation for $\psi$. 
As is well known, classical statistical mechanics, where a 
probability distribution in phase space evolves over time according
to the Liouville equation, exhibits many similarities with
quantum statistical mechanics, where a Wigner function evolves in the 
same phase space \cite{Wigner,Kim&Noz}. In this phase space picture, 
there are 
two familiar differences between 
the classical and
quantum-mechanical descriptions:
\begin{enumerate}
\item
The quantum time evolution is slightly different, except for 
Hamiltonians that are of quadratic or lower degree in positions
and momenta.
\item
Certain classically allowed sets of initial data are forbidden
by the Heisenberg uncertainty principle.
\end{enumerate}
Although of great relevance in many microphysical 
problems \cite{Zurek&Paz}, the
first of these two differences is irrelevant for our present 
discussion, as the correction terms are small, and moreover 
have the effect of making the phase-space time-evolution even 
more non-linear and chaotic. Thus for our purposes, we can get qualitatively
correct results by ignoring this first difference, and simply 
replacing the quantum treatment by the treatment in terms of 
classical random fields which has become standard in cosmology,
and for which we already know the answer.
The second difference is the crucial one here, since as we will see, 
it implies that
the initial $\psi$ cannot simply correspond to the trivial solution
of $\delta$ being constant with vanishing time derivatives.
For a single particle, the uncertainty principle implies that 
$\Delta x\,\Delta p\ge\hbar/2$, 
which means that the compulsory $p\neq 0$ components will cause a localized
wave packet to gradually ``spread out". For the familiar textbook problem 
of a pencil perfectly balanced on its tip, the uncertainty principle analogously
forbids the trivial classical solution where the pencil remains balanced forever.
Our gravitational instability example is analogous, except that we are dealing 
with a quantum field $\delta(\vr)$ with an infinite number of
degrees of freedom. The uncertainly principle 
simply bounds from below the product of $\Delta\delta(\vr)$ with the 
spread in the corresponding conjugate momentum, thus forbidding the trivial 
classical solution with no fluctuations (where $\delta$ retains a constant 
value throughout space with no spread, corresponding to the balanced-pencil solution).

In summary, the initial field $\delta(\vr)$ must be in a 
superposition of having several different values at each point, 
and will therefore change due to gravity.
We thus draw the following conclusion:
since any one initial field configuration with small fluctuations 
will eventually evolve into a messy and lumpy mass distribution,
the initial state $\psi$, being a superposition of such states,
will evolve into a superposition of messy and lumpy states.

\Subsection{3.3. Symmetry is Still not Broken}

\noindent
It should be stressed that although a type of symmetry 
breaking has occurred for each individual element of the superposition
(approximate translational invariance has been broken),
the wavefunction $\psi$ retains all the symmetry that it originally
had, since gravity itself is translationally and rotationally
invariant. If the the initial $\psi$ had exact translational invariance
(which means that the one-dimensional probability distribution 
for $\delta(\vr)$, the two-dimensional probability distribution 
for $[\delta(\vr_1),\delta(\vr_2)]$, {\etc} were all invariant 
under a shifting of the coordinate system), then the resulting 
wavefunction today will have this same property. 
Where is all the irregular gravitational clumping hidden?
The answer is to be found in all sorts of intricate long-range correlations
that did not exist in the initial $\psi$. For an extreme example, 
suppose that the time-evolution regrouped all the mass into homogeneous
lumps the size of Earth. Then the correlation
between $\delta(\vr_1)$ and $\delta(\vr_2)$ would be almost perfect
if $\vr_1$ and $\vr_2$ were separated by one meter, since if 
$\vr_1$ is in a planet, $\vr_2$ probably is as well.
$\psi$ would nonetheless be translationally invariant, 
corresponding to superpositions of planets at different locations, with
all locations given equal weight.
In summary, non-linear dynamics with chaotic behavior will 
create a type of hidden complexity within the wavefunction $\psi$, but not 
itself cause symmetry breaking. In the following section we will 
argue that the symmetry breaking is apparent, 
as well as discuss why it appears to be broken 
in a way that favors ``classical" states.


\goodbreak\bl\bl\noindent{\bf 4. UNIVERSALLY VALID QUANTUM MECHANICS}
\Subsection{4.1. A Brief Review of the No Collapse Interpretation}

\noindent
Although quantum mechanics has proven to be an astonishingly 
successful theory by any standards, the issue of how it is to be interpreted has 
caused heated debate ever since it was created some 70 years ago. 
At the center of the controversy stands the so called collapse postulate, 
according 
to which the 
time-evolution of the wavefunction is occasionally {\it not}
governed by the Schr\"odinger equation, but 
described by a discontinuous random jump, a so-called ``collapse".
This postulate was introduced to explain the apparently random outcomes
of certain quantum measurements, but is completely 
unnecessary according to the Everett
interpretation \citerange{Everett1}{DeWitt},
also known
as the ``many worlds interpretation".
The debate continues about the underlying 
philosophical and ontological issues 
(see, {\eg}, \cite{Peres&Zurek,Kent}), 
but these are not crucial to the argument in this paper.
All that is needed here is the assumption that quantum 
mechanics is universally valid:
\begin{itemize}
\item 
{\it
The time-evolution of any isolated system is 
governed by the\\
 Schr\"odinger equation.
}
\end{itemize}
This is what Albert refers to as the ``raw theory". 
Thus all isolated systems, not merely microscopic ones, 
are assumed to obey quantum mechanics, 
be they planets, human beings or the entire universe.
In other words, there is no 
wave function collapse built into the theory 
itself. Rather, this phenomenon is assumed to
arise automatically on a phenomenological level. 
Everett argued that a generic observer will subjectively 
experience an {\it apparent} wavefunction collapse consistent 
with the familiar quantum probabilities.
The basic idea is illustrated by the following simple example.
Suppose Ang\'elica prepares $n$ different spin$-{1\over 2}$ 
silver atoms to have spin up
in the $x$-direction, and then proceeds to measure their spin in 
the $z$-direction with a Stern-Gerlach apparatus. 
She writes down the results in the form of binary digits, 
$0$ denoting ``down" and $1$ denoting ``up", with a decimal point
in front of it all, so that the outcome of the entire set
of measurements can be 
codified as a real number $x$ on the interval $[0,1]$. For instance, 
if $n=4$, the outcome  ``up-down-down-up" would correspond to 
$x=0.1001$. In the limit where $n\to\infty$, there is thus a 
one-to-one correspondence between outcomes and real numbers
on $[0,1]$. 
In the Everett interpretation, the final state for the system 
consisting of both Ang\'elica, her laboratory and her silver atoms
is thus a superposition of orthogonal states that we can 
label by numbers $x$, such
that state $x$ corresponds to the atoms having precisely 
the spins given by the decimals of $x$ and Ang\'elica having experienced
exactly these measurement results.
The key point realized by Everett is that almost all Ang\'elicas in 
this final superposition will have measured spin up $50\%$ of 
the time, in exactly the same sense that almost all real numbers on
$[0,1]$ have $50\%$ of their binary digits equal to one
(the set of those that do not has Lebesgue measure zero).
Thus almost all Ang\'elicas in the final superposition will 
find the outcome of their experiment in agreement with
the standard quantum probabilities.
The generalization to finite $n$ and more generic probabilities than $50\%$
is also presented by Everett, and in every case, he finds that almost all observers
in the final superposition will find their 
experimental results in agreement with 
the probabilities that would arise from the collapse hypothesis.
Certain aspects of Everett's arguments are still controversial, but as mentioned, 
these technical details are of no importance for the main argument 
in the present paper.

\Subsectionn{4.2.}{How Universal Quantum Mechanics Causes}{Apparent Complexity}

\noindent
The above example brings us back to the argument of Section 2.1,
except that we are now discussing reality\footnote{\basel
In this paper, we use the word ``reality" to denote that 
which is described by $\psi$. In other words, we think of all elements of 
the a superposition as being equally real, bearing in mind that 
in any particular element of the superposition, there may be an 
observer who subjectively feels that merely that element is real.
(This is of course merely an issue of semantics, so all that matters
is that we clearly state how we choose to use various words,
deferring to philosophers the question of which usage is preferable.)}
rather than physical models. 
To rephrase the argument in quantum language, there is relatively little 
algorithmic information in the final state $\psi$. 
To describe any one particular
outcome in the grand superposition, however, would require
$n$ bits of information, and we can for arguments sake choose $n$ 
to be enormous. Thus although reality ($\psi$) remained as simple
as it was before the experiment, a generic Ang\'elica
in the final superposition will be staring at a piece of paper with 
a long and seemingly random string zeroes and ones and consider 
this to be lots of information.
Thus {\it subjectively}, as experienced by a self-aware subset of
$\psi$ (Ang\'elica), the complexity and algorithmic information
of the world has increased, although {\it objectively}, 
when viewing the mathematical object $\psi$ from outside, the algorithmic
information of $\psi$ stayed constant.

This is precisely what happens in the example of cosmological
structure formation discussed in the previous section.
Even if nonlinear dynamics produces a wavefunction $\psi$ where
a planet is in a superposition of being centered in two different places,  
separated by one meter, this is of course not what a self-aware subset
of $\psi$ would experience. Rather, after the observer had interacted
with the planet (by absorbing electromagnetic radiation 
reflected from it, say), $\psi$ would for all practical 
purposes be a superposition of two orthogonal states, each corresponding
to the planet having a definite location and the observer finding
it in that same location. In other words, 
when we gaze into the night sky and see stars
all over the place (in definite positions, but positions that appear
somewhat random), this is exactly what we would have expected to see. 
However, in the picture put forward here, the seemingly large information
content in these stellar positions (information that has by now filled
many volumes of astronomy publications) is merely information
in the eyes of the beholder.
It is not real information in the
algorithmic sense, and would never enter in a compact description
of $\psi$ itself. 

\Subsection{4.3. The Role of Decoherence}

\noindent
The no collapse interpretation of quantum mechanics leaves a crucial
question unanswered \cite{Ben-Dov,Kent,SquiresPosition}: 
why is the position operator $\opr$ so special?
In other words, why is it that we tend to find macroscopic 
objects in (approximate) eigenstates
of $\opr$ in particular, even though there are infinitely
many other Hermitean operators (for instance $\opr + \op$) 
whose eigenstates involve macro-superpositions (such as the
above-mentioned 
superposition of a planet being in two macroscopically
distinct places)?
A breakthrough on this issue was not made until more than a decade later
\cite{Zeh}, and it took another decade until the question was  
definitely answered \cite{Zurek81,Zurek82}.
The answer, usually referred to as decoherence (see \cite{Zurek91} 
for a recent review),
lies in the fact that macroscopic systems are virtually
impossible to isolate from their surroundings.
It can be shown that the net effect of this 
inevitable interaction is to suppress off-diagonal elements
of the spatial density matrix, which means that to a subjective observer,
a macroscopic object will 
for all practical purposes 
appear to have a definite position.\footnote{\basel
The 
deeper question of why the calculations always end up
favoring the position operator is related to the 
unexplained empirical fact that the Lagrangians
of nature seem to be spatially local (as opposed to being local in 
for instance Fourier space).}
It is straightforward to quantify the decoherence effect caused by 
various interactions that we know to occur
\cite{Joos&Zeh,Collapse},
and one finds that even such relatively weak effects as scattering
of microwave background photons and (on earth) solar neutrinos will
keep an object the size of a bowling ball localized to within 
a small fraction of an atomic radius.

Decoherence is quite crucial to the main argument in this paper.
If Ang\'elica had measured the spin of her silver atoms in the 
$x$-direction instead, she would simply have measured ``up" 
every time. The quantum creation of apparent complexity that was
described above of course requires that we are measuring an
observable that the quantum system is not already in an eigenstate of.
Thanks to decoherence, our classical world is very position centered,
so our minds are constantly keeping track of where things are located.
As we saw in the previous section, nonlinear gravitational evolution
will automatically convert a translationally invariant initial state
into a state where macroscopic objects are {\it not} in position
eigenstates, but rather in a superposition of being in many different
places. This is why we should expect to find apparent complexity 
in the sky. 
Electromagnetic, weak and strong forces of course
all produce non-linear time-evolution as well, so we have no right to be
surprised by finding apparently 
complex structures on more everyday scales either
--- after all, non-linear dynamics is so ubiquitous that
it has been jocularly compared to 
``the zoology of non-elephants".

\Section{5. DISCUSSION}

\noindent
We have argued that although the universe appears to contain 
an enormous, perhaps even infinite, amount of information, 
this impression
is nonetheless consistent with the assumption that its algorithmic
information content is quite small, even if no new physics whatsoever
is invoked. In short, reality could in fact be 
much simpler than it appears, with its apparent 
plethora or complex structures such as trees, fountains and rabbits.
The enormous discrepancy between apparent and actual
information content would have been caused by an interplay of several
well-known effects:
\begin{itemize}
\item
Non-linear dynamics evolved quantum states that were initially only
microscopically different into quantum states that were macroscopically
different.
\item
Well-known decoherence effects ensured that only  
macroscopically
\linebreak 
``classical" states could be perceived by
self-aware subsets of the universe (such as us), so for all 
practical purposes, the wavefunction of the universe
is in a superposition of a such states.
\item
Although the total wavefunction still contains as little algorithmic
information as it did in the beginning (assuming that the laws of physics
that govern its time evolution are known), 
any one element of this superposition (which is all that 
we can subjectively experience, because of decoherence) 
will display the enormous complexity brought about by the
chaotic non-linear evolution.
\end{itemize}
Thus even if both the laws of physics and the initial conditions exhibit 
so much symmetry that they are easy to describe, we would not
expect most of these symmetries to be manifested in what we see when we 
open our eyes. For instance, we would not expect the matter distribution
around us to be translationally invariant even if the initial wave function 
was.

\Subsection{5.1. Why Universal Quantum Mechanics is Needed}

\noindent
It should be emphasized that if the wavefunction really
collapses, then the main 
argument of this paper does as well. In other words, in those 
interpretations of quantum mechanics where the Schr\"odinger evolution
is replaced by some random process while Ang\'elica measures her 
spins, leaving a final wavefunction $\psi$ corresponding to merely one 
definite sequence of outcomes, the algorithmic information of 
$\psi$ will in fact increase.
The reason that it is no longer conserved is of course the randomness
itself, since causality has been lost and the initial data no 
longer suffice to specify the final wavefunction. Rather, the randomness will
increase the algorithmic information with one bit per 
spin measurement.

\Subsection{5.2. Why Chaos is Needed}

\noindent
Why does the above argument fail in the absence of 
non-linear\footnote{\basel
Here and throughout we refer to 
a system as linear if its 
Hamiltonian is at most quadratic. This corresponds to the classical 
equations of motion in phase space being linear. 
Thus the phase-space distributions 
at two different times are simply related by a linear canonical transformation.
For such quantum systems, the Wigner functions at two different times are
related by this same linear canonical transformation.
This type of linearity should not be confused with that of the 
Schr\"odinger equation, which is of course linear (in $\psi$)
however nonlinear the Hamiltonian may be.
}
dynamics?
Roughly speaking, the apparent symmetry breaking caused by quantum mechanics 
and decoherence applied only to 
{\it macro-superpositions}, whereas the only superpositions
that are required to exist initially by the Heisenberg uncertainty principle is
{\it micro-superpositions}. Thus for our argument to work, we need some process to 
evolve the initial micro-superpositions into macro-superpositions.
In our example of cosmic structure formation, we needed a mechanism that converted the 
microscopic density fluctuations of the early universe into planets {\etc} that were in
superpositions of being in many macroscopically separated positions at once.
Converting micro into macro is of course the very essence of chaotic dynamics, 
which exponentially amplifies microscopic differences in initial data
as time passes.

\Subsection{5.3. What about Computability?}

\noindent
In the definition of algorithmic information, the universal 
computer is of course required to produce the desired output 
in a finite amount of time. 
In other words, the current wavefunction of the universe should 
be computable from the initial state and the equation of motion,
to any prescribed accuracy,  in 
a finite number of program steps,  
As long as the equations of motion are non-chaotic, this will 
be achievable by integrating the partial-differential 
equations of motion numerically, 
using some time-step $\Delta t$ that 
is chosen small enough to fulfill the
accuracy requirement.
If the equations of motion are chaotic, however, 
this simple-minded approach
of course fails in practice, since the computed final 
state would be significantly wrong unless an absurdly high numerical 
precision was used, the number of significant digits 
required growing exponentially with 
the duration between the initial and final time.
Fortunately, the relevant quantity in our case is the 
universal wave-function $\psi$, whose evolution is non-chaotic, since
it is governed
by the linear 
Schr\"odinger equation. Thus ironically, although the 
chaotic nature of the corresponding classical dynamics was crucial
to a different aspect of our argument, the non-chaotic nature of 
pure quantum evolution is important as well.

In summary, whereas quantum states themselves evolve non-chaotically, 
the reduced density matrices describing subsystems {\bf do} evolve 
chaotically, and both of these formal facts are essential to our
argument.

\Subsection{5.4. What about the Initial State?}

\noindent
It should be emphasized that although we assumed that the initial state 
of the universe had quite a small algorithmic information content,
our argument about apparent complexity holds more generally,
in the sense that an observer will always perceive the universe 
to be more complicated than it actually is. 
Roughly speaking, the observer's description 
(as given by the appropriate reduced density matrix) will
inherit all the complexity of the big bang, and in addition 
include substantial amounts of extra ``illusory" complexity.

\Subsection{5.5. What about Entropy?}

\noindent
The subtle relationship between entropy and various measures of information
has been discussed by many authors (see {\eg} 
\cite{ZurekAlgo1,ZurekAlgo2,Lloyd} for reviews), often in connections with
Maxwell's infamous demon.
It has been emphasized \cite{Lloyd&Bagels} that neither a state of 
minimum disorder (such as a perfect crystal) nor maximum disorder 
(such as an ideal gas in thermal equilibrium) 
corresponds to what we with a course-grained description 
would describe as complex.
Rather, the hallmark of perceived complexity is the so called 
``depth", which often reduces to the difference between the 
coarse-grained and fine-grained entropies \cite{Lloyd&Bagels}.
This provides an alternative way summarizing the arguments above: 
the fine-grained entropy of the universe remains constant,
but non-linear dynamics and the apparent symmetry-breaking brought 
about by quantum mechanics and decoherence increase the coarse-grained 
entropy that we perceive, thus increasing the depth, the 
apparent complexity.

In addition to these processes, 
the course-grained entropy of any open system will of course increase
in accordance with the 
second law of thermodynamics, even in classical physics,
as the system gradually becomes correlated with ever more 
distant parts of the universe --- see {\eg} \cite{Lloyd} 
for a recent review, including a discussion of when the 
entropy can decrease.
However, such an entropy increase merely reflects an increase in our 
ignorance about the system in question, not an increase in 
the complexity of the system.
Consider a visible particle suspended in a static liquid, 
undergoing Brownian motion. If we observe it at some definite position, 
and then look away from our microscope, 
our subjective probability distribution for the particle position will 
widen over time as $\sqrt{t}$. Yet when we eventually look again, 
thereby reducing our ignorance and thus reducing the entropy, the system 
we see appears neither more nor less complex than it did to begin with.
 
\Subsection{5.6. Would it be Testable?}

\noindent
Is this entire discussion merely a useless metaphysical digression without
any experimentably testable consequences? 
Not necessarily. Ever since the big bang theory became firmly established, 
the problem of understanding the initial conditions has aroused interest. 
For instance, one of the main goals of modern cosmology is to accurately measure 
the power spectrum of density fluctuations. By comparing the fluctuations 
measured by 
large galaxy redshift surveys to the fluctuations measured in the cosmic 
microwave background radiation, strong constraints can be placed on models
for the formation of large scale structure in the universe. Most of the currently 
popular models 
(cold dark matter, mixed dark matter, baryonic dark matter, {\etc}) 
simply predict how a ``primordial" power spectrum would evolve due to 
various microphysical effects, which means that 
all these models must be supplemented
with an assumption about the form of this primordial spectrum.
One popular prediction of what the primordial power spectrum should be
is the inflationary scenario (see {\eg} \cite{Kolb&Turner} for a review), 
but many other suggestions
have been put forward as well, including ones appealing to 
quantum gravity, {\eg} \cite{Hawking,Souradeep}.
The point is that the initial conditions prevailing shortly
after the big bang are not necessarily in the realm of 
metaphysics, and may turn out to have measurable consequences.
If this turns out to be the case, then it will perhaps be possible for 
us to use our knowledge of the mechanisms described above to
make inferences about the initial conditions from observational data,
and if these initial conditions are indeed simple enough to have a small
information content, we may even be able to determine them with reasonable 
accuracy.

\Subsection{5.7. Would it be Good News or Bad News?}

\noindent
A pessimistic way of summarizing the scenario put forward in this paper 
would be as yet another step in the human humiliation that began with the
Copernican revolution. After being dethroned from our position in the center
of the universe half a millennium ago, forced to accept that we are orbiting the sun
rather than vice versa,
our human egos have received a series of devastating blows. 
We have now been demoted further, to orbiting a rather average star a rather average
distance from the center of a rather average galaxy, our sun being but one star 
amongst $10^{11}$ others
in but one galaxy amongst at least that many more.
Alas, the humiliation shows no signs of abating.
In the cold dark matter scenario and variants thereof, the bulk of the 
matter in the universe is conjectured to be non-baryonic, snobbishly 
refusing to interact with us much at all, 
leaving us as some sort of ignored fringe minority.
Our human self-confidence has been deflated on the mental side as well.
Darwin told us that we were
descended from amoebas and differed from other organisms not in a qualitative way
but merely in a quantitative way. 
Freud claimed that we did not even understand ourselves
all that well, and Skinner argued that we had no free will,
being more like organic computers that could be programmed by suitable 
conditioning. Last year, Garri Kasparov was beaten by a chess computer.
Now we are arguing that even the intricate structures that are the subjects of
our thoughts, dreams and efforts in life, everything from our loved ones to
our parking tickets, are perhaps more aptly described as illusions,
as manifestations of the fact that our minds experience the grandeur of reality
merely from an extremely limited frog perspective. 
More precisely, we are arguing that it is the grandeur itself that 
is merely apparent, since the 
complete description of the wavefunction of the universe is 
orders of magnitude simpler than what we see when we open our eyes.
In short, reality would be much more banal than it appears to be.

However, we can just as well choose to summarize our 
discussion on an optimistic note.
First of all, once we have abandoned the notion 
of us humans playing a central role
in the universe, and forced ourselves to find a {\it raison d'\^etre} elsewhere,
yet another step away from the limelight clearly makes little difference. 
The everyday reality to which we wake up in the mornings will of course
remain the same whether we change our ontological views 
or not --- no amount of 
philosophizing will make those parking tickets go away.
If we thus restrict our value judgements to empirical considerations, 
the picture put forward here would have quite positive implications for our 
ability to do science in the future.
The information content of the world around us naively appears to completely
dwarf the sort of information quantities that we humans can store and process. 
However, if the algorithmic information content of the universe is indeed quite small,
then perhaps we need not despair when faced with the vast and complex
world around us.
Rather, this would raise the slight hope that we humans 
may one day be able to understand it.
\bl\bl



\def\refname{\basel{\bf REFERENCES}}

\end{document}